
%
%
\magnification=1200
\pretolerance=10000
\baselineskip=18pt
\voffset -1 true cm
\centerline{\bf COMBUSTION GASDYNAMICS IN PROTO-NEUTRON STARS}
\bigskip
\bigskip
\bigskip
\centerline {J.E.Horvath}

\bigskip
\centerline{\it Instituto Astron\^omico e Geof\'\i sico, Universidade
de S\~ao Paulo}
\centerline{\it Av.M.St\'efano 4200 (04301-904) S\~ao Paulo SP BRASIL}
\bigskip
\bigskip
\bigskip
\bigskip
\bigskip
\noindent
{\bf ABSTRACT}
\bigskip
The possible conversion of neutron matter to strange matter and its
relevance to astrophysical problems are discussed. Particular attention
is paid to the {\it form} in
which this hypothetical "burning" process should be propagated
in dense matter and to the {\it moment} when it should start.
\vskip 8 true cm
\centerline {Talk given at the {\it VI J.A.Swieca Summer School on Nuclear}}
\centerline {{\it Physics} - Campos do Jord\~ao - Brazil - February 1993}
\vfill\eject
\noindent
{\bf 1. INTRODUCTION}
\bigskip
\noindent
Since first stated by Witten [1] the hypothesis of {\it strange matter}
(hereafter SQM) being the true ground state of cold hadronic matter has
attracted considerable attention [2]. Among the several aspects studied
within the strange matter hypothesis we may mention the problem of the
conversion of (cold) neutron matter to strange matter in astrophysical
objects. This has been first studied in [3,4] and refined in [5,6] and all
these works have {\it assumed} that the process is analogous to an ordinary
(slow) combustion. This is of course very reasonable in a first approximation,
but unfortunately things seem not to be that simple. Not only the actual way
in which the conversion develops is subject to considerable doubt, but
the {\it moment} when it can start is also controversial. This point turns out
to be very important because it can alter dramatically the fate of the host
star. We shall attempt to summarize in this work what is known (and unknown)
about this problem, which appears to us far from being closed.
\bigskip
\noindent
{\bf 2. HOW AND WHEN DOES THE BURNING START ?}
\bigskip
\noindent
A variety of mechanisms for the conversion to start have been discussed in [7],
tentatively classified as "primary" (in which SQM is produced somehow inside
the dense environment) and "secondary" (SQM or a large amount of energy to
produce it are seeded from the outside of the star). Clearly, the secondary
mechanisms, even though perfectly viable, are much difficult to assess and
and we shall not discuss them here.

Perhaps the most likely primary mechanism discussed in the literature is the
conversion {\it via} two-flavor quark matter. The scenario relies on the
possibility of deconfining the light quarks from ordinary nucleons as the
stellar matter is compressed/cooled in a supernova aftermath. This is sketched
in Fig.1 ; the transition point is achieved at a density $n_{c}$ and the
quark region relaxes to the SQM state on a timescale typical of weak
interaction $\tau_{w}$ lowering its energy per baryon number unit. From the
"macroscopic" point of view, and since $\tau_{w} \ll \tau_{dynamical}$ the
process may be modelled as isocore. The released energy difference builds up
an overpressure $\Delta P$ whose whose value can be found for a given model
calculation. In [8] we have employed a free massless quark gas $+$ bag
pressure and the Bethe-Johnson I  [9] equation of state for neutron matter to
calculate $\Delta P \; \simeq \; 70^{+}_{-}10 \; MeV \; fm^{-3}$, depending
somewhat on the parameters. This result shows that the overpressure is huge
and its decay may be very important dynamically (see below). Note that, if
at all, this process must occurr shortly after the gravitational collapse
itself since a significative compression is achieved on a timescale
$\tau \; \simeq \; 1 \; s$ [10].
\vskip 7 true cm
\noindent
{\bf Fig.1.} A possible route for SQM appearance viewed in the free-energy
{\it vs.} density plane. Deconfinement is achieved at a value $n_{c}$ (dashed
line) and relaxation to SQM is further achieved on a weak-interaction timescale
(pictured with a wavy line).
\bigskip
If SQM is not produced as the result of two-flavor quark matter "relaxation",
it is possible that it could arise because of thermal nucleation of SQM
bubbles inside hot nuclear matter. The idea is that if a single bubble
of the fluctuation distribution grows beyond a critical size [11] it will
convert the sourroundings afterwards. A simple calculation using a kinetic
 Fokker-Planck-Zel'dovich approach has been presented in [12]. The growth of
the bubbles has been described by the expression

$$ {{\partial f}\over {\partial t}} = - {{\partial \xi}\over {\partial r}}
\eqno (1) $$
\bigskip
where $ f(r,t) $ is the time-dependent size distribution of bubbles and
$\xi$ is the nucleation rate. The appropiate solution for this problem may
be written as

$$ \xi = 2.2 \, 10^{-2} \; {(T/ \sigma)^{1\over 2}} \; N_{qc}^{3\over 4} \;
\tau_{w}^{-1} \; exp(-3.1 \; {N_{qc}^{1\over 2}} \; \sigma /T) \eqno(2) $$
\bigskip
with $N_{qc}$ the number of quarks of the critical size bubble ($\leq \, 300$),
$\tau_{w}$ a weak-interaction timescale and $\sigma$ the surface tension
coefficient. A lower bound to the temperature at which at least one bubble
larger than the critical size will nucleate can be found by imposing

$$ \xi \; \Delta t \; V_{nuc} \; \geq \; 1 \eqno (3) $$
\bigskip
where $\Delta t$ is the time available for prompt nucleation ($\sim \; 1 \; s$)
 and $V_{nuc} \; \sim \; (1 \; km)^{3}$ is the volume in which the nucleation
can take place. As $\sigma \; \leq \; (70 \; MeV)^{3}$ from detailed
calculations [13] and fittings to phenomenological models [14], the condition
(3) can be re-written equivalently as

$$ T \geq 2 \; MeV \eqno(4) $$
\bigskip
This bound can be certainly satisfied in the first seconds of the neutron
star life but {\it not} after that since the initial cooling phase is dominated
by neutrino emission, which is in turn a strong function of $T$. Therefore, we
again conclude that the conversion is likely to start immediately after the
gravitational collapse itself. The same conclusion holds if the conversion is
triggered by "dormant" strangelets [15] which become active as soon as a full
neutronization of the core is possible ($\tau \; \sim \; 1 \; s$).
\bigskip
\noindent
{\bf 3. WHICH IS THE PREFERRED COMBUSTION MODE ?}
\bigskip
\noindent
As we have discussed in section 2, a short timescale $\sim \; 1 \; s$ for SQM
to be "operative" inside a proto-neutron star appears to be a general feature
of any triggering mechanism. This is, of course, not enough to address the
aftermath, since the form in which the conversion front propagates remains
unclear. For a given chemical potential difference $\Delta \mu$, the conversion
$n \rightarrow  \; SQM$ releases energy much in the same way as a chemical
reaction does. Therefore, the well-known theory of combustions [16] can be
applied to study the process. Physically, steady state combustions can occur
in two modes : deflagrations and detonations. Deflagrations (or {\it slow}
combustions) are subsonic burnings in which quasi-equilibrium among the burnt
and unburnt fluids holds. Detonations (or {\it rapid} combustions) are
supersonic and can be described as a shock wave followed by the combustion
front.

Assuming that a slow combustion occurrs, the velocity of the flame has been
calculated to be $\upsilon_{s} \, \sim$ few $km/s$ at most [3,4]. These works
 have focused mainly on the reaction kinetics of the conversion, but the
actual boundary conditions in a star and the gas motion should play a role
as well. To study these effects, a linear stability analysis of this
propagation mode has been performed [17] in which the stability condition
was found to be

$$ j^{4} \; < \; 4 \; \sigma \; g \; \rho^{2}_{1} \rho^{2}_{2} \; { 1 \over
{(\rho_{2} - \rho_{1})}}  \eqno(5) $$
\bigskip
\noindent
with $j$ the mass flux onto the flame, $\sigma$ the surface tension, $g$ the
local gravitational acceleration and $\rho_{1} , \rho_{2}$ the fluid densities
of the unburnt and burnt sides respectively. On the other hand, by definition
[16]

$$  j^{2} = {{(P_{2} - P_{1}) \rho_{1} \rho_{2}} \over {(\rho_{2} - \rho_{1})}}
  \eqno(6) $$
\bigskip
\noindent
from which either $P_{2} > P_{1}$ and $\rho_{2} > \rho_{1}$ (detonations) or
$P_{2} < P_{1}$ and $\rho_{2} < \rho_{1}$ (deflagrations). The important point
is that if we impose the latter conditions [18], then eq.(5) can never be
satisfied. We conclude that deflagrations are always Rayleigh-Taylor unstable
in dense matter for any perturbation wavenumber at the linear level.

These results indicate that, unless the instability can be controlled in some
way (e.g. by non-linear effects), the flame should become turbulent and
accelerate. Laboratory observations show that, in such a case, the foldings
of the front increase, thus increasing the area and the local burning rate.
The heat transfer mechanism switches from thermal conduction to turbulent
convection, which gives a much larger velocity $\upsilon_{s}$. An estimate of
the latter can be given in terms of a lenght scale, taken to be the radius of
the spherical front $r$ (the only available lenght of the problem) divided by
the timescale for the instability to grow $\tau_{RT}$. Given that $\tau_{RT}$
must necessarily be some multiple of
the dynamical time $(G \rho)^{-1/2}$ we find

$$ \upsilon_{turb} \; \simeq \; {r \over {\tau_{RT}}} = \; 10^{9} \;
{\bigl( {r \over {1 \; km}} \bigr)} \; cm \; s^{-1} \eqno(7) $$
\bigskip
\noindent
which probably contains an uncertainty of a factor of 10 in either direction.
The point here is that not only $\upsilon_{turb} \gg \upsilon_{s}$ but also it
is likely to be comparable to the speed of sound $c_{s}$ at some finite radius.
However, since the turbulent convection burning must also be subsonic, its
description is also contained in the lower branch of the Hugoniot adiabatic.
Note also that, due to the supranuclear density values of the stellar matter,
$\tau_{RT} \; \simeq \; 1 \, ms$ and thus the acceleration process will be very
fast. Moreover, a further change of the combustion mode is to be expected, as
we shall now discuss.

We have previously noted that boundary conditions must be taken into account
for a reliable and consistent description of the flow. As is well-known [16],
these boundary conditions (analogous to the "closed pipe" ones) force the
existence of a shock wave propagating ahead of the flame. The strenght of the
shock is uniquely determined by the velocity of the flame and the state of the
unburnt material. Now, if the flame accelerates the shock becomes stronger,
 which in turn means that the neutrons are more and more pre-heated and
pre-compressed before meeting the flame. Such pre-heating and pre-compression
effects on the matter entering the combustion zone help to further
accelerate the burning rate and establishes a positive
feedback loop [19].
\midinsert
\vskip 5 true cm
\noindent
{\bf Fig.2.} {\it a)} Laminar initial stage of the deflagration,
arrows indicate the velocity of the flame and pre-compression shock
{\it b)} folding of the flame due to instability grow {\it c)} fully developed
turbulent flame (note the acceleration of the pre-compression shock
$-broken \; arrow-$)
\endinsert
\bigskip
The crucial point is therefore to establish which is the velocity of
the flame for which the shock is strong enough to "ignite" the neutron matter
by itself. This type of calculations have been longly performed for
laboratory gases (for planar and spherical geometries) and it has been
repeatdely found [19] that for $\upsilon_{turb} \; \simeq \; 0.1 \; c_{s}$ the
so-called auto-ignition condition is met and thus the only possible
propagation is the one corresponding to the Chapman-Jouget point on the
Hugoniot adiabat [16].

Very recent work [20] on this problem has revealed another
important feature of the
$n \rightarrow SQM$ combustion : it has been found that deflagrations are
{\it not} allowed unless $ \rho_{1} < \, 1.63 \, \rho_{o}$ ( $\rho_{o}$ being
the nuclear saturation density). The reason for this is simply that for
higher densities the velocity of the flame becomes larger than the velocity
of the pre-compression shock. Physically this means that for this regime the
flame and the shock coalesce in a single region, a feature that characterizes
a detonation wave [19]. However, given the short timescale for the flame to
become turbulent, it is very likely that the detonation develops a width
$\delta \; \sim \; \tau_{w} c_{s} \; \leq $ 1 $m$.
So, even though the kind of combustion is limited by
the the weak interactions rate (producing $s$ quarks), the heat transport
mechanism must be recognized as a fundamental ingredient in the problem.
It appears from the above results that we should consider {\it ab initio}
a (turbulent) detonation mode.

To conclude this section two remarks are in order. First, some doubt has been
expressed about the ability of "slow" weak interactions to sustain an energy
release fast enough for a detonation to propagate. We note that, even for the
longest reasonable timescale of $\sim \; 10^{-8} \; s$ the front thickness is
always negligible with respect to the dimension of the burning region
($\sim$ km) and, much more important, with respect to the lenghtscale of
significative variations of the pressure, etc. estimated as
$L = {\bigl( {\partial ln P/ \partial r)} \bigr)}^{-1} $. In fact
$\delta \; \ll \; L$ is a sufficient condition for
the stability of a propagating detonation front [21], or in other words,
"slow" weak reactions are fast enough to sustain a detonation unless (say)
$L \; < \; 10-100 \; m$ . To be
more quantitative, however, a complete numerical simulation, including the
(strong) dependence of the weak interactions rate with $T$
[6] should be carefully
performed. The second remark is that it is by no means necessary that the
burning passes through an initial slow combustion stage and then accelerates.
On the contrary, a "direct" initiation is more likely if SQM is produced by the
 decay of a two-flavor plasma in bulk (section 2). In this case, the
overpressure $\Delta P$ decays into a shock wave with strenght
$\delta P \; \simeq \; \Delta P /2$, which
should be enough to ignite the matter by itself (see [22] for a through
discussion of this and related points).
\bigskip
\noindent
{\bf 4. CONCLUSIONS}
\bigskip
\noindent
We have discussed in this talk several topics related to the possible
appearance and propagation of SQM inside dense neutron matter. We have
argued that
\medskip
\noindent
a) SQM is expected to form/begin to "operate" after $\sim$ 1 $s$ following the
collapse at most, and we are not aware of any mechanism to postpone the
conversion until, say, ages $\sim \, yr$ (so that {\it all} neutron stars must
be strange stars [1,2,7,12,18]).
\medskip
\noindent
b) Even if a deflagration may develope first [3,4], it should become
turbulent and switch to a
detonation shortly after. Alternatively, a detonation may be directly
initiated.
\medskip
The conclusions (to which a factor $0 \; \leq f \leq \; 1$ gauging the
uncertainties of the problem may be added) point to the need for a better
understanding of compact star interiors, specially for the
role of few-quark states. Preliminary steps have been taken in [23]. It is
also very important to improve the schematic model in which the sudden
energy release from a detonation affects the supernovae events (hopefully in
a positive way !) [8,24]. The conceptual similarity of the latter with the
"delayed detonation model" for type I supernovae [21] is quite remarkable
and may merit a closer look. Note, however, that even if a detonation is
{\it not} formed after all, the very presence of a strong turbulent flame
propagating at high (but subsonic) velocity may be more than enough to add
substantial energy to a stalled prompt shock.
\bigskip
\noindent {\bf 5. ACKNOWLEDGEMENTS}
\bigskip
\noindent
It is a pleasure to acknowledge the financial support of the Organizing
 Committee of the J.A.Swieca School
and the CNPq, Brasil for a Research Fellowship award.
\bigskip
\noindent {\bf 6. REFERENCES}
\bigskip
\noindent
[1]  E.Witten, {\it Phys.Rev.D}{\bf 30}, 272 (1984).

\noindent
[2]  See the {\it Proceedings of the Workshop on Strange Quark Matter in
Physics and Astrophysics}. {\it Nuc. Phys. B Proc. Supp.}{\bf 24} (1991).

\noindent
[3]  G.Baym,E.Kolb,L.Mc Lerran,T.Walker and R.L.Jaffe,
{\it Phys. Lett. B}{\bf 160}, 181 (1985).

\noindent
[4]  A.V.Olinto, {\it Phys. Lett. B}{\bf 192}, 71 (1987).

\noindent
[5]  M.L.Olesen and J.Madsen, Ref.2.

\noindent
[6]  H.Heiselberg, G.Baym and C.J.Pethick, Ref.2.

\noindent
[7]  C.Alcock and A.V.Olinto, {\it Ann. Rev. Nuc. Part. Sci.}{\bf 38},
161 (1988).

\noindent
[8]  O.G.Benvenuto, J.E.Horvath and H.Vucetich, {\it Int. Jour. Mod. Phys. A}
{\bf 4}, 257 (1989).

\noindent
[9]  H.Bethe and M.Johnson, {\it Nuc. Phys. A}{\bf 230}, 1 (1974).

\noindent
[10] A.Burrows and J.Lattimer, {\it Astrophys.J.}{\bf 307}, 178 (1986).

\noindent
[11] E.M.Lifshitz and I.Pitaevskii, {\it Physical Kinetics}
(Pergamon, London, 1980).

\noindent
[12] J.E.Horvath, O.G.Benvenuto and H.Vucetich, {\it Phys. Rev. D}{\bf 45},
3865 (1992).

\noindent
[13] M.L.Olesen and J.Madsen, {\it Phys. Rev. D}{\bf 47}, 2313 (1993) ;
 M.S.Berger, {\it Phys. Rev. D}{\bf 40}, 2128 (1989) ;
M.S.Berger and R.L.Jaffe, {\it Phys. Rev. D}{\bf 35}, 213 (1987).

\noindent
[14] E.Farhi and R.L.Jaffe, {\it Phys. Rev. D}{\bf 30}, 2379 (1984).

\noindent
[15] O.G.Benvenuto and J.E.Horvath, {\it Mod. Phys. Lett. A}{\bf 4}, 1085
(1989).

\noindent
[16] L.D.Landau and E.M.Lifshitz, {\it Fluid Mechanics}
(Pergamon, New York, 1980).

\noindent
[17] J.E.Horvath and O.G.Benvenuto, {\it Phys. Lett. B}{\bf 213}, 516 (1988).

\noindent
[18] Note also that the "naive" expectation $\rho_{2} > \rho_{1}$ (SQM should
be denser than the nuclear one) {\it forces} a detonating conversion.

\noindent
[19] J.H.S.Lee and I.O.Moen, {\it Prog. Energy Combust. Sci.}{\bf 6},
359 (1980).

\noindent
[20] G.Lugones and H.Vucetich, private communication.

\noindent
[21] A.M.Klokhov, {\it Astron. Astrophys.}{\bf 246}, 383 (1991).

\noindent
[22] T.J.Mazurek and J.C.Wheeler, {\it Fund. Cosm. Phys.}{\bf 5}, 193 (1980).

\noindent
[23] O.G.Benvenuto, J.E.Horvath and H.Vucetich, {\it Phys. Rev. Lett.}{\bf 64},
713 (1990).

\noindent
[24] O.G.Benvenuto and J.E.Horvath, {\it Phys. Rev. Lett.}{\bf 63}, 716 (1989).

\bye